# Performance Evaluation of WiMAX (802.16) Using Different Encoding Schemes

Waqar Asif, Muhammad Bilal Qasim, Syed Musa Raza Tirmzi, Usman Muhammad Khan
Department of Electrical Engineering
COMSATS Institute of Information Technology, Islamabad, Pakistan
waqar.asif@gmail.com

*Abstract*— **This paper deals with the performance of Worldwide Interoperability for Microwave Access (WiMAX), when we enhance its physical layer attributes with help of different encoding techniques. For this evaluation Space Time Block Codes (STBC) and Turbo codes are separately introduced into the architecture of WiMAX that works on adaptive modulation technique.**

*Keywords*- **WiMAX, Adaptive Modulation, Orthogonal Frequency Division Multiplexing, Guard Band, Inverse Fourier Transform, Space Time Block Codes, Maximum Likelihood Detector, Turbo Codes, Interleaver, Log Map Decoder.**

## I. INTRODUCTION

WiMAX stands for Worldwide Interoperability for Microwave Access. WiMAX is based on wireless Metropolitan Area Network (MAN) standards developed by IEEE 802.16 group. It operates in the licensed exempt and licensed spectrum between 2-11 GHz and 10-66 GHz frequency ranges respectively [1]. It was developed to connect the Internet and to provide a last mile wireless extension to cable and DSL broadband access. IEEE 802.16 provides up to 50 km of linear service area range and allows users connectivity without a direct Line of Sight (LOS) to a Base Station (BS). The technology also provides shared data rates up to 70 Mbps, which is enough bandwidth to simultaneously support many users.

In Space Time Block Codes (STBC), data stream is encoded in blocks and distributed among time and spaced antennas. At the transmitter side this encoded data is transmitted along multiple antennas whereas, at the receiver side multiple antennas receive multiple copies of the same signal and then extract the best possible out of it. It provides significant increase in throughput and range without any increase in the overall bandwidth and transmits power expenditure. It also increases the spectral efficiency (number of information bits per hertz of bandwidth) of wireless system by using multiple antennas that are separated in space and time [2].

Turbo codes are the best approximation of the Shannon limit. In Turbo codes at the transmitter side a single bit is encoded into a combination of bits depending on the architecture of the encoder. When this encoded data is transmitted over the channel the probability of error is reduced to a great extent. When the data reaches the receiver side the data is decoded back into the original bits that are understandable by the receiver.

## II. WiMAX

A number of industry standards govern the design and performance of wireless broadband equipment. The standards that chiefly concern wireless broadband are 802.16 and its derivative 802.16a, both of which were developed by the Institute of Electrical and Electronic Engineers (IEEE), a major industry standards body headquartered in the United States.

### A. *Physical Layers of WiMAX*

The IEEE 802.16 standard supports multiple physical specifications due to its modular nature. The first version of the standard only supported single carrier modulation and after the passage of time and as technology grew; OFDM and scalable OFDMA were also used but only for operating in the Non Line of Sight (NLOS) environment and were meant to provide mobility. The standards were then further enhanced to work in the lower frequency range of 2-11GHz along with the previous 10-66GHz band.

*1) Orthogonal Division Multiplexing (OFDM):* The idea of OFDM comes from Multi Carrier Modulation (MCM) transmission technique. MCM is based on the division of input bit stream into several parallel bit streams and then using them to modulate several sub carriers as shown in Fig 1. Guard band is introduced in between each subcarrier so that they do not overlap with each other. This guard band also supports the band pass filter on the receiver side in identifying individual subcarriers. OFDM is a special form of spectrally efficient MCM technique. It differs from its predecessors in the way that it uses orthogonal subcarriers which also eliminate the use of a band pass filter from the receiver side. The orthogonal nature of the subcarriers also removes the Inter Carrier Interference (ICI) which was a great problem previously. The guard band previously used is also removed in OFDM hence concluding in the reduction of bandwidth usage as shown in the Fig. 2.

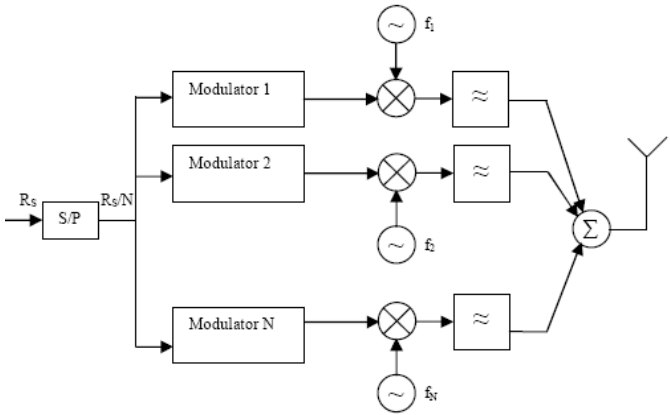

Fig. 1. OFDM Transmitter

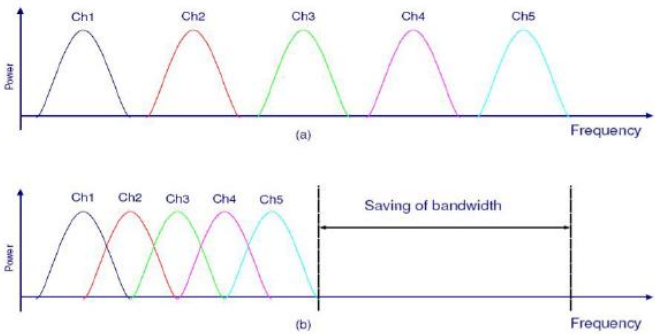

Fig. 2. OFDM Symbols

*2) Adaptive Modulation and Coding:* AMC allows WiMAX systems to select the most appropriate Modulation and Coding Scheme (MCS) depending on the propagation conditions of the communication channel, e.g., if the propagation conditions are good, a higher order modulation scheme with a lower coding redundancy is used which also increase the data rate of the transmission, while on the other hand during a signal fade, the system selects a modulation scheme of a lower order to maintain both the connection quality and the link stability without an increase in the signal power [3]. For this purpose WiMAX uses four modulation schemes that are:

- Binary Phase Shift Keying (BPSK)
- Quadrature Phase Shift Keying (QPSK)
- 16- Quadrature Amplitude Modulation (16- QAM)
- 64- Quadrature Amplitude Modulation (64- QAM)

### III. SYSTEM MODEL

The first step on the physical layer of WiMAX is the decision of the modulation scheme that has to be used as shown in Fig 3. Initially the transmitter transmits the bits using 16-QAM modulation. The modulated data is then mapped on to orthogonal channels using IFFT. In this mapped data, to avoid inter-carrier interference a guard band (of 1/8 of the total number of bits) is added. The data is then transmitted over the channel and at the receiver end the guard band removed and the data decoded back into the original form. The receiver calculates the error in the received data and informs the transmitter of this. The transmitter uses this information to change its modulation scheme as per the requirement.

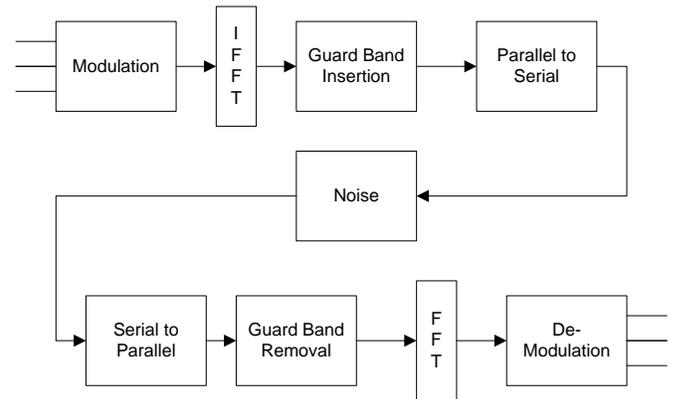

Fig. 3. WiMAX Module

*A. Introduction of STBC in WiMAX System*

Fat the transmitter the STBC Encoder takes the data from the guard band insertion block as shown in Fig 4 and transmits that data over two spaced antennas. Different symbols are simultaneously transmitted over these antennas to reduce noise interference. The receiver after receiving the signal retrieves the bits using Maximum Likelihood decoding algorithm and passes the data to the guard band removal block. As shown in Fig 4.

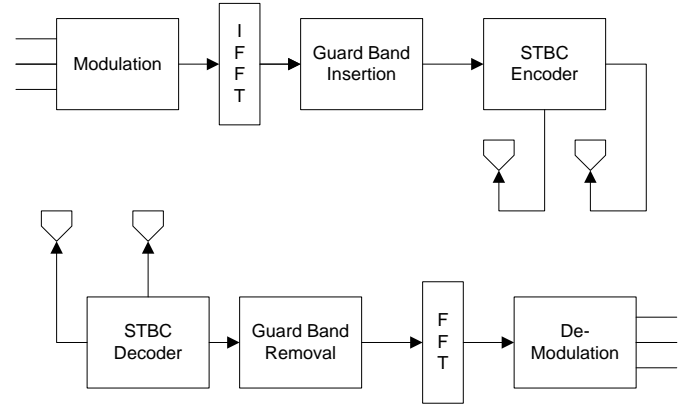

Fig. 4. WiMAX Module with STBC

*1) Encoding and Transmission sequence:* For achieving diversity in time and space symbols are transmitted according to the following matrix

Table I.
THE ENCODING AND TRANSMISSION SEQUENCE FOR THE TWO-BRANCH TRANSMIT DIVERSITY SCHEME

|          | Antenna 0 | Antenna 1 |
|----------|-----------|-----------|
| Time t   | $s_0$     | $s_1$     |
| Time t+T | $-s_1^*$  | $s_0^*$   |

For Antenna 0 at time t the complex multiplicand distortion can be modeled as $h_0(t)$ and for Antenna1 it can be modeled as $h_1(t)$. Now if we assume that fading is constant across two symbols then we can write [4].

$$h_0(t) = h_0(t + T) = \alpha_0 e^{j\theta}$$
$$h_1(t) = h_1(t + T) = \alpha_1 e^{j\theta}$$
$$h_2(t) = h_2(t + T) = \alpha_2 e^{j\theta}$$
$$h_3(t) = h_3(t + T) = \alpha_3 e^{j\theta} \quad (1)$$

This can be expressed in matrix form as

Table II.
THE DEFINITION OF CHANNELS BETWEEN THE TRANSMIT AND RECEIVE ANTENNAS

|  | Rx Antenna 0 | Rx Antenna 1 |
|---|---|---|
| Tx Antenna 0 | $h_0$ | $h_2$ |
| Tx Antenna 1 | $h_1$ | $h_3$ |

The received signals can be expressed as

$$r_0 = h_0 s_0 + h_1 s_1 + n_0$$
$$r_1 = -h_0 s_1^* + h_1 s_0^* + n_1$$
$$r_2 = h_2 s_0 + h_3 s_1 + n_2$$
$$r_3 = -h_2 s_1^* + h_3 s_0^* + n_3 \quad (2)$$

here $n_0$, $n_1$, $n_2$ and $n_3$ are complex random variable representing receiver thermal noise and interference [4].

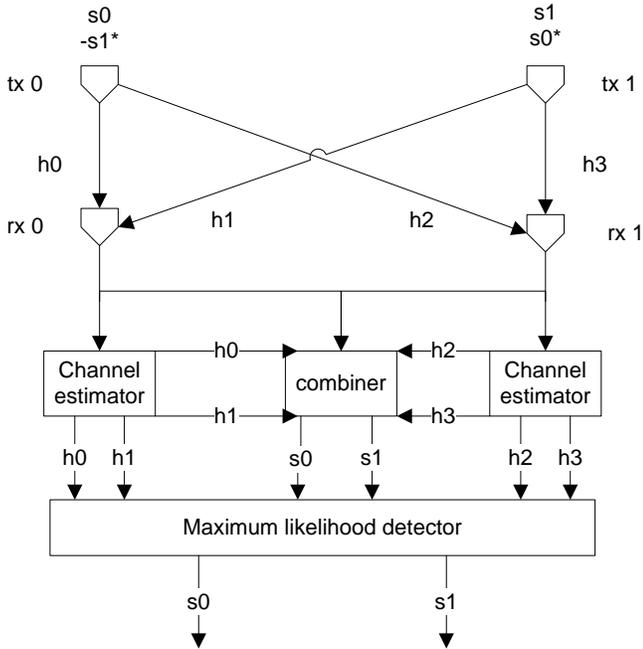

Fig. 5. Two-branch transmit diversity scheme with two receivers

*2) The combining Scheme:* The combiner of Fig 6 builds the following two combined signals and sends them to the maximum likelihood detector

$$s_0 = h_0^* r_0 + h_1 r_1^* + h_2^* r_2 + h_3 r_3^*$$
$$s_1 = h_1^* r_0 - h_0 r_1^* + h_3^* r_2 - h_2 r_3^* \quad (3)$$

*3) The Maximum Likelihood decision Rule:* The maximum likelihood uses the decision rule given below to determine the original transmitted signals. It chooses $s_i$ if:

$$(\alpha_0^2 + \alpha_1^2 + \alpha_2^2 + \alpha_3^2 - 1)|s_i|^2 + d^2(s_0, s_i) \leq (\alpha_0^2 + \alpha_1^2 + \alpha_2^2 + \alpha_3^2 - 1)|s_k|^2 + d^2(s_0, s_k) \quad (4)$$

After this the symbols obtained are sent to the guard band removal section of OFDM and the normal procedure of WiMAX module continues its work.

*B. Introduction of Turbo Codes in WiMAX System*

Turbo Encoder uses the bits passed on to it by the MAC layer of WiMAX and with help of Recursive Systematic Convolution Encoder encoded the bits and passes them on to the modulation scheme block. At the receiver the Log MAP decoder takes bits from the demodulation block and passes this data on to the MAC layer after decoding it [5]. This is shown in Fig 6.

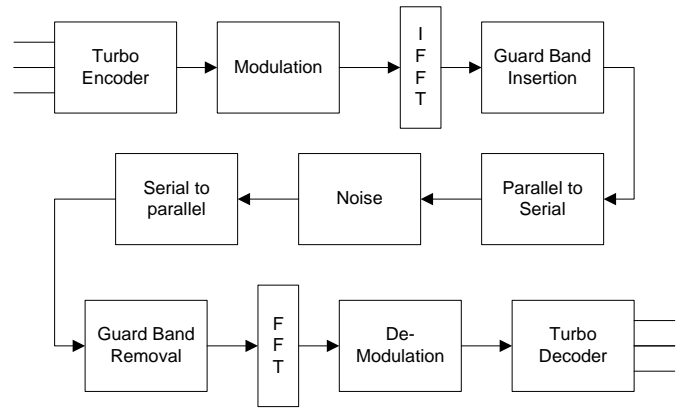

Fig. 6. WiMAX Module with Turbo Codes

*1) Recursive Systematic Convolutional Encoder (RSC):* This type of encoder works on a bit by bit basis. For every input bit it generates a *parity bit* depending on the structure of the encoder and outputs the same input bit at the output known as the *systematic bit*. The encoder is implemented using Linear Feedback Shift Register (LFSR). These registers are the main reason why we call it Recursive process. The output is feed back to the input and every new output is dependent on the previous input to the encoder.

In Fig 6 the D denotes the registers which are initially at a known state of 00. After a packet is encoded the registers can be in any one of the four possible states 00,01,10,11 depending on the previous input. These states cause problem for the next packet. To remove this problem and to bring the registers back into a known state of 00, *Memory Flushing or Trellis Termination* is done. For memory flushing we pad *Trail Bits* at the end of each packet depending on the state of the registers.

*2) Interleaver:* Interleaver is an essential part of the encode encoder. It is used to change the sequence of the bits in

a frame. In the encoding stage the data is interleaved before the data is feed to the second encoder.

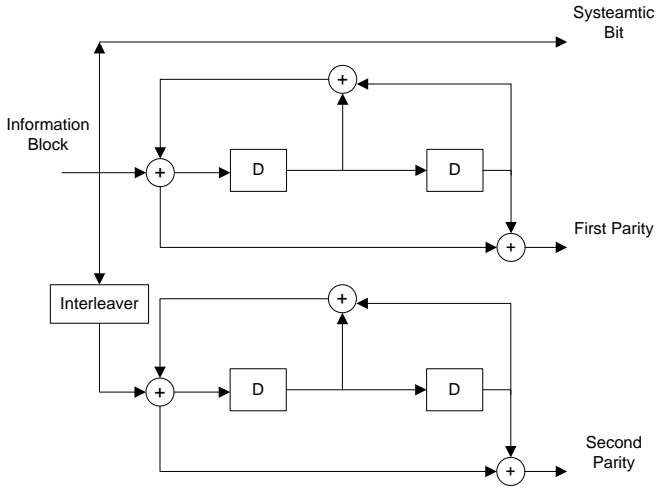

Fig. 7. Recursive Systematic Convolutional Encoder

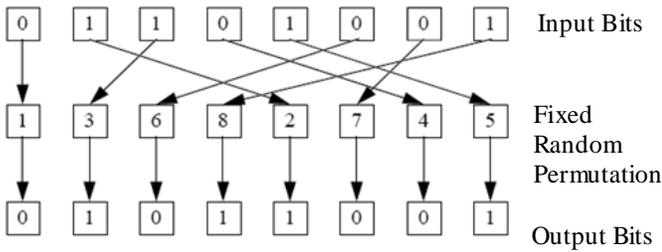

Fig. 8. Random Interleaver

The interleaver of Fig 8 uses a fixed random permutation and maps the input according to that permutation order. Here the length of the input is denoted by L the greater the size of L the better will be the performance Fig 8 shows that if we take L=8 then the interleaver inputs a sequence [0 1 1 0 1 0 0 1] and outputs [0 1 0 1 1 0 0 1].

*3) Log Map Decoder:* To achieve a soft decision through turbo decoding, two Log Map Decoders are used together which work iteratively on a symbol by symbol decoding method.

*a) Symbol By Symbol decoding*: For symbol by symbol decoding procedure we break our observations into three main parts. The first part denotes the past observations before time k, the second part denotes the present observations at time k and the third and the last part denotes the future observations after time k .i.e. [6]

$\alpha_k(s)$ is the joint conditional probability of the past observation before time k and

$$\tilde{\alpha}_k(s) = \frac{\sum_{s'} \tilde{\alpha}_{k-1}(s') \gamma_k(s', s)}{\sum_{s} \sum_{s'} \tilde{\alpha}_{k-1}(s') \gamma_k(s', s)} \quad (5)$$

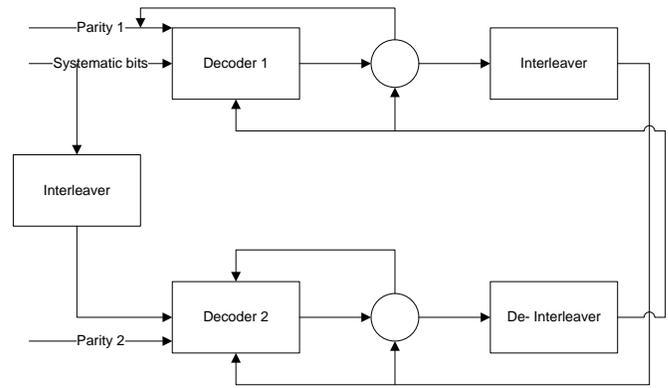

Fig. 9. Turbo decoder

Here $\sum_{s'/s_k=s}()$ gives us the over all summation of the possible states $s_{k-1}$ whose end at a state where $s_k$=s. The initial condition that is defined as

$$\tilde{\alpha}_0(s) = \begin{cases} 1 & \text{if } s = 1 \\ 0 & \text{otherwise} \end{cases} \quad (6)$$

means that the start of the trellis states for the encoder is at state 1.

$\beta_{k-1}(s)$ is the joint conditional probability of the future observations after time k that is at k-1.this is also computed recursively as follows

$$\tilde{\beta}_{k-1}(s') = \frac{\sum_{s} \tilde{\beta}_k(s) \gamma_k(s', s)}{\sum_{s} \sum_{s'} \tilde{\alpha}_{k-2}(s') \gamma_{k-1}(s', s)} \quad (7)$$

Here $\sum_{s'/s_k=s}()$ is the summation over all the possible states $s_{k-1}$ whose ending state is $s_k$=s. The initial condition is

$$\tilde{\beta}_N(s) = \begin{cases} 1 & \text{if } s = 1 \\ 0 & \text{otherwise} \end{cases} \quad (8)$$

which means that the encoder Trellis states always start at state 1.

$\gamma_k(s)$ is the joint conditional probability of the present observation at time k and

$$\gamma(s', s) \propto \exp\left[\frac{1}{2} \cdot L^e(c_k^1) \cdot c_k^1 + Lc \cdot \frac{1}{2} \cdot y_k^{1,s} \cdot c_k^1\right]$$
$$\exp\left[\sum_{i=2}^{q}\left(Lc \cdot \frac{1}{2} \cdot y_k^{i,p} \cdot c_k^i\right)\right] \quad (9)$$

These three probabilities are combined to compute the symbol using.

$$L(u_k) = \log[\frac{\sum_{u+}\tilde{\alpha}_{k-1}(s')\gamma_k(s',s)\tilde{\beta}_k(s)}{\sum_{u-}\tilde{\alpha}_{k-1}(s')\gamma_k(s',s)\tilde{\beta}_k(s)}] \qquad (10)$$

Both the decoders compute the symbol using the symbol by symbol decoding method and continue this process on an iterative basis till the time they get to a similar result. When they both reach at the same result the decoding process is stopped and the result extracted. This process repeats itself for every upcoming packet.

## III. SIMULATION RESULTS

The Simulation model was implemented in Matlab 7. The simulation was run for a packet size of $10^6$, and for better approximation this packet was transmitted $10^2$ times.

Fig. 10. shows the result obtained for the performance of WiMAX, using adaptive modulation technique.

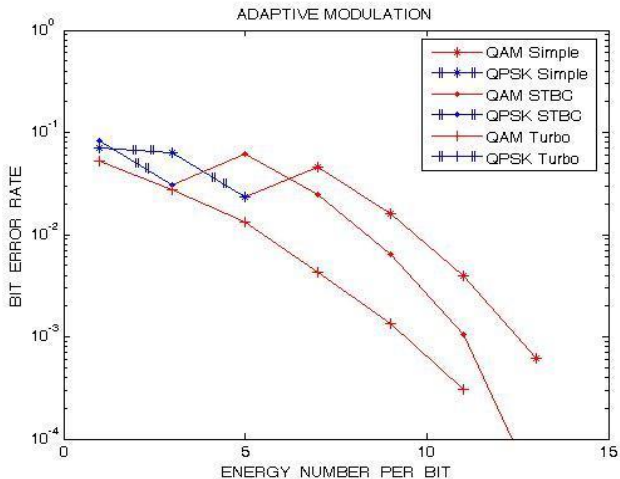

Fig. 10. Performance Graph of WiMAX

Initially the packet was sent over simple WiMAX architecture and the graph marked with '*' achieved. It shows initially QPSK modulation is used till an energy number per bit of five and after that it starts working at QAM. It means that the atmosphere is too noisy that's why initially QPSK is used to successfully transmit symbols but as soon as the BER reaches a value below our defined threshold it shifts to a higher modulation scheme i.e. QAM.

In the similar condition when we introduce STBC in the architecture of WiMAX we have a graph marked with '.'. In this graph the modulation scheme shifts from QPSK to QAM at an energy number per bit of three. This shows that for lower values of energy number per bit the observed BER is high but as the energy number increases the BER starts decreasing at a increased rate.

The third graph marked with '+' is also of the similar condition as were for simple WiMAX architecture. The difference here is that there is no need to transmit our symbols over QPSK. The whole data is transmitted using QAM modulation. This denotes that when we introduce Turbo Coding into our architecture the effect of noise is greatly reduced.

## IV. CONCLUSION

It is concluded that when we introduce Turbo Codes into WiMAX we have a very improved BER, but the Trellis Termination bits, the tail bits and the parity bits make up an over head that is not feasible for a wireless link like WiMAX. On the other hand by the introduction of STBC we have an improved BER compared to simple WiMAX architecture with a very low overhead. So we suggest that to improve the performance of WiMAX, STBC should be introduced in it.

## V. REFRENCE

[1] Syed Ahson and Mohammad Ilyas, WiMAX Technologies, Performance Analysis and QoS, CRC Press, Taylor & Francis Group, 2008.
[2] ]Hamid Jafarkhani, " A Quasi- Orthogonal Space time Block Code," IEEE Trans Commun, vol 49,pp 1-4, Jan 2001
[3] Dania Marabissi, Daniele Tarchi, Romano Fantacci and Francesco balleri, "Efficient Adaptive modulation and Coding techniques for WiMAX systems," ICC proceedings 2008.
[4] Siavash M. Alamouti, " A Simple Transmit Diversity Technique for Wireless Communications," IEEE journals, vol 16, no. 8, October 1998.
[5] C. Berrou, A. Glavieux, and P. Thitimajshima, "Near Shannon limit error-correcting coding and decoding: Turbo-codes,: in Proc. ICC,pp.1064-1070,1993.
[6] S. Benedetto, D. Divsalar, G. Montorsi, and F. Pollara," Soft-Output Decoding Algorithms in Iterative Decoding of Turbo Codes", TDA progress report 42-124, Feburary 15,1996
[7] Chien- Ming Wu. Ming-Der Shieh, Chien-Hsing Wu, ying-Tsung Hwang, Jun-Hong Chen, and Hsin-Fu Lo," VLSI Architecture Exploration for Slifting Window Log-Map Decoders",ISACS 2004